# Dark Pulse Emission of a Fiber Laser

**H. Zhang[£], D. Y. Tang*, L. M. Zhao, and X. Wu**

School of Electrical and Electronic Engineering,

Nanyang Technological University, Singapore 639798

*: edytang@ntu.edu.sg, corresponding author.

£: zhan0174@ntu.edu.sg, zhanghanchn@hotmail.com

Personal website: http://www3.ntu.edu.sg/home2006/ZHAN0174/

We report on the dark pulse emission of an all-normal dispersion erbium-doped fiber laser with a polarizer in cavity. We found experimentally that apart from the bright pulse emission, under appropriate conditions the fiber laser could also emit single or multiple dark pulses. Based on numerical simulations we interpret the dark pulse formation in the laser as a result of dark soliton shaping.

Pulsed operation of lasers is an extensively studied topic of laser physics [1]. It is well-known that a laser can be forced to emit optical pulses through either the laser Q-switching or mode locking techniques. While Q-switching of a laser generates optical pulses with nanosecond duration, mode locking of a laser can produce much narrower pulses, which are normally in the picosecond or femtosecond levels depending on the laser gain bandwidth and the specific method used. Both the Q-switched and the mode locked lasers are nowadays widely used in various fields such as the industrial material processing, medical treatment, military and scientific researches.

However, so far all pulsed lasers can only emit bright pulses. It would be interesting to know whether a laser could also emit dark pulses. Here dark pulses are referred to as a train of intensity dips in the intensity of a continuous wave (CW) background of the laser emission. We have known the answer. In this paper we report on a dark pulse emission fiber laser. We show experimentally that under appropriate operation conditions, an all-normal dispersion cavity fiber laser can emit a train of single or multiple dark pulses. Dark pulse formation in the fiber laser was also numerical investigated. Based on results of numerical simulations we interpret the dark pulse formation as a result of the dark soliton shaping in the laser.

Our fiber laser is schematically shown in Fig. 1. Its cavity is made of all-normal dispersion fibers. A piece of 5.0 m, 2880 ppm Erbium-doped fiber (EDF) with group velocity dispersion (GVD) of -32 (ps/nm)/km was used as the gain medium. The rest of fibers used have altogether a length of 157.6 m, and they are all dispersion compensation fiber (DCF) with GVD of -4 (ps/nm)/km. All the passive components (WDM, Coupler, Isolator) were made of the DCF. A polarization dependent isolator was used in the cavity to force the unidirectional operation of the ring, and an in-line polarization controller was inserted in the cavity to fine-tune the linear birefringence of the cavity. A 50% fiber coupler was used to output the signal. The laser was pumped by a high power Fiber Raman Laser source (KPS-BT2-RFL-1480-60-FA) of wavelength 1480 nm, and the maximum pump power can be as high as 5W. An optical spectrum analyzer (Ando AQ-6315B) and a 350MHZ oscilloscope (Agilen 54641A) together with a 2GHZ photo-detector were simultaneously used to monitor the laser emission.

The laser cavity has a typical configuration as that of a fiber laser that uses the nonlinear polarization rotation (NPR) technique for mode locking [2, 3]. Indeed, under appropriate linear cavity phase delay bias (LCPDB) setting, self-started mode-locking could be achieved in the laser. Consequently either the gain-guided solitons or the flat-top dissipative solitons could be obtained [4, 5]. Apart from the bright soliton operations, experimentally we have further identified another regime of laser operation, where a novel form of dark pulse emission as shown in Fig. 2a was firstly revealed. Fig. 2a (upper) shows a case of the laser emission where a single dark pulse was circulating in the cavity with the fundamental cavity repetition rate of 1.23MHz (inset of Fig. 2b). On the oscilloscope trace the dark pulse is represented as a narrow intensity dip in the strong CW laser emission background. The full width at the half minimum of the dark pulse is narrower than 500ps, which is limited by the resolution of our detection system. Unfortunately, due to the low repetition rate of the dark pulses, their exact pulse width cannot be measured with the conventional autocorrelation technique, but a cross-correlation measurement is required. In Fig. 2b we have shown the optical spectrum of the dark pulses. For the purpose of comparison, the CW spectrum of the laser emission is also shown in the same figure. Obviously, under the dark pulse emission the spectral bandwidth of the laser emission became broader. Based on the 3dB spectral bandwidth and assuming that the dark pulses have a transform-limited hyperbolic-tangent profile, we estimate that their pulse width is about 8 ps.

To obtain the above dark pulse emission, the LCPDB must be shifted away from the mode locking regime [3]. This was experimentally done by carefully adjusting the polarization controller. In the non-mode-locking regime strong stable CW laser emission was obtained, whose strength increased with the pump power. In our experiment as the launch pump power increased to ~2W, the CW laser intensity suddenly became strongly fluctuated. Carefully checking the CW laser intensity fluctuation, it turned out that clusters of dark pulses were formed in the background of the CW intensity. The random movement and mutual interactions of the dark pulses caused the strong intensity fluctuation of the CW laser emission beam intensity. Experimentally, we found that through carefully adjusting the pump strength and orientations of the PC, the number of dark pulses could be significantly reduced. Eventually a state of stable single dark pulse emission as shown in Fig 2a (upper) was obtained.

Unlike the single bright pulse emission of the laser, the single dark pulse emission state was difficult to be maintained for long time. Probably due to the laser noise and/or weak environmental perturbations, new dark pulses always automatically appeared in the cavity, leading to states of multiple dark-pulse operation, an example is shown in Fig. 2a (down trace). Different from the multiple gain-guided soliton operation of the laser, where all the solitons have the same pulse height and energy, known as the soliton energy quantization [6], obviously each of the multiple dark pulses has different shallowness, indicating that neither their energy nor their darkness is the same.

By removing 150m DCF from the cavity, we had also experimentally studied dark pulse emission under short cavity length. Dark pulses could still be obtained, however, their appearance required much higher pump intensity, and a stable single dark pulse state was difficult to achieve. In addition, by replacing the polarization dependent isolator with a polarization independent one, we could still obtain the dark pulse emission of the fiber laser. However, there is no NPR induced cavity feedback effect in the laser. Again in the laser it is hard to obtain a stable single dark pulse emission state. These experimental results suggest that the dark pulse formation could be an intrinsic feature of the all-normal dispersion cavity fiber lasers, and the NPR induced cavity feedback could have played an import role on the stability of the dark pulses in the laser.

To confirm our experimental observations, we have numerically simulated the dark pulse formation in our laser based on a round-trip model as described in [3]. To make the simulations comparable with our experiment, we used the following parameters: the orientation of the intra-cavity polarizer to the fiber fast birefringent axis $\Phi=0.125\pi$; nonlinear fiber coefficient $\gamma=3\ W^{-1}km^{-1}$; erbium fiber gain bandwidth $\Omega g$ =24nm; fiber dispersions $D''_{EDF}$= -32 (ps/nm) /km, $D''_{DCF}$= -4(ps/nm) /km and $D'''$=0.1($ps^2$/nm)/km; cavity length L=5.0$m_{EDF}$+157.6$m_{DCF}$=162.6m; cavity birefringence $L/L_b$=0.1 and the nonlinear gain saturation energy $P_{sat}$=50 pJ.

We have always started our simulations with an arbitrary weak dip input. It was found that when the LCPDB was set in the laser mode locking regime, the self-started mode locking occurred and a bright pulse was always obtained. We have therefore focused our simulations on the LCPDB values in the non-mode-locking regime. With low laser gain, independent of the value of LCPDB no dark pulse could be obtained. However, under strong laser gain it was found that if the LCPDB was so selected that the laser cavity provided a large negative cavity feedback (reverse saturable absorption), a dark pulse could automatically formed in the laser. Fig. 3(a) shows a dark pulse numerically obtained under a gain coefficient of 900 $km^{-1}$ and a linear cavity phase delay bias of 1.0 π. Fig. 3b shows the intensity and phase profiles of the dark pulse. The pulse intensity has a hyperbolic-tangent shape and a brutal phase jump close to π was observed at the minimum of the pulse, suggesting that it is a dark soliton [7-14]. Fig. 3c shows the spectrum of the calculated dark pulse. It closely resembles the spectra of the experimentally observed dark pulses, as shown in Fig. 2. The soliton feature of the pulse is also reflected by the appearance of spectral sidebands in its spectrum. Nevertheless, due to the unavoidable experimental noise these spectral sidebands could not be detected experimentally. Numerically further increasing the laser gain, the CW background level was boosted, while the absolute depth of the dark soliton kept the same, leading to a decrease on the soliton darkness. As $G > 4000\ km^{-1}$, only a noisy CW background was observed.

Based on the numerical simulations we interpret the formation of dark pulses as a result of dark soliton shaping in the laser. Due to the normal dispersion of the cavity fibers, dark solitons are automatically formed under strong CW operation. As a dark soliton can be created by an arbitrary initial small dip on a CW background [7], and in the practical laser noise is unavoidable, many dark solitons are always initially formed in the laser. Formation of each individual dark soliton is uncorrelated. Depending on the initial "seed dips" the dark soliton have different darkness. Therefore, to obtain a stable single dark pulse train of the laser emission, a certain competition mechanism among the solitons is required, which explains why a long cavity is beneficial for obtaining a stable single dark soliton in the laser. Finally, we point out that as the formed dark pulses have only a narrow spectral bandwidth no gain filter effects occur. Therefore, the effect of gain in the laser is purely to provide a strong CW and compensate the laser losses.

In conclusion, we have experimentally demonstrated a dark pulse emission fiber laser. It was shown that under strong CW operation, dark pulses could be automatically formed in a fiber laser made of all-normal dispersion fibers. Through introducing a strong negative cavity feedback mechanism, stable single pulse could be selected. Eventually a train of dark pulses at the fundamental cavity repetition rate was obtained. To our knowledge, this is the first demonstration of such a dark pulse emission laser.

## Figure captions:

Fig.1: Schematic of the fiber laser. WDM: wavelength division multiplexer. EDF: erbium doped fiber. DCF: dispersion compensation fiber. PDI: polarization dependent isolator. PC: polarization controller.

Fig. 2: Dark pulse emission of the laser. (a) Oscilloscope traces, upper: single dark pulse emission; down: multiple dark pulse emission. (b) Optical spectra of the laser emissions. Inset: RF spectrum of the single dark pulse emission.

Fig. 3: A dark pulse state numerically calculated. (a) Evolution with cavity roundtrips. (b) Intensity and phase profile. (c) Optical spectrum.

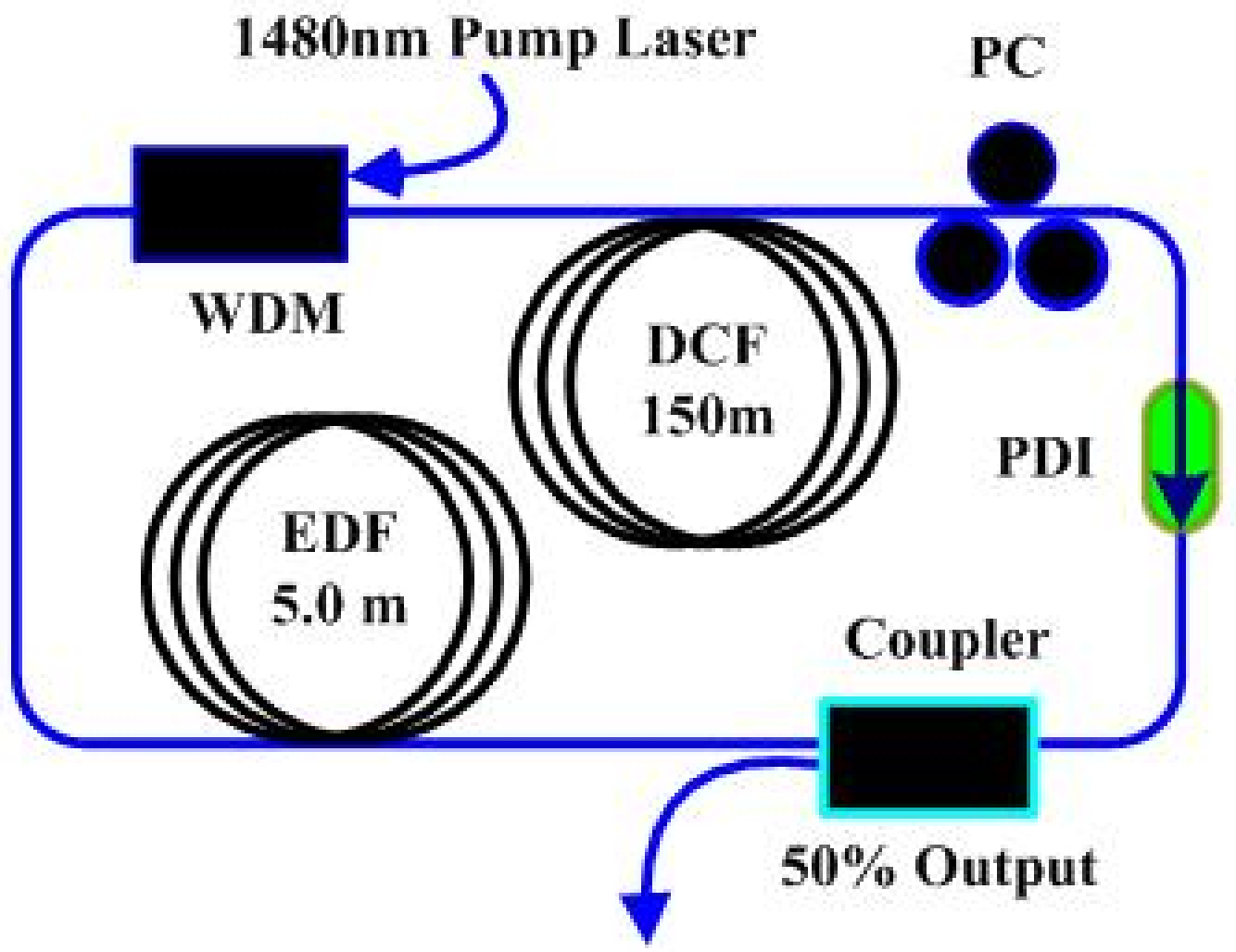

Fig.1 H. Zhang et al

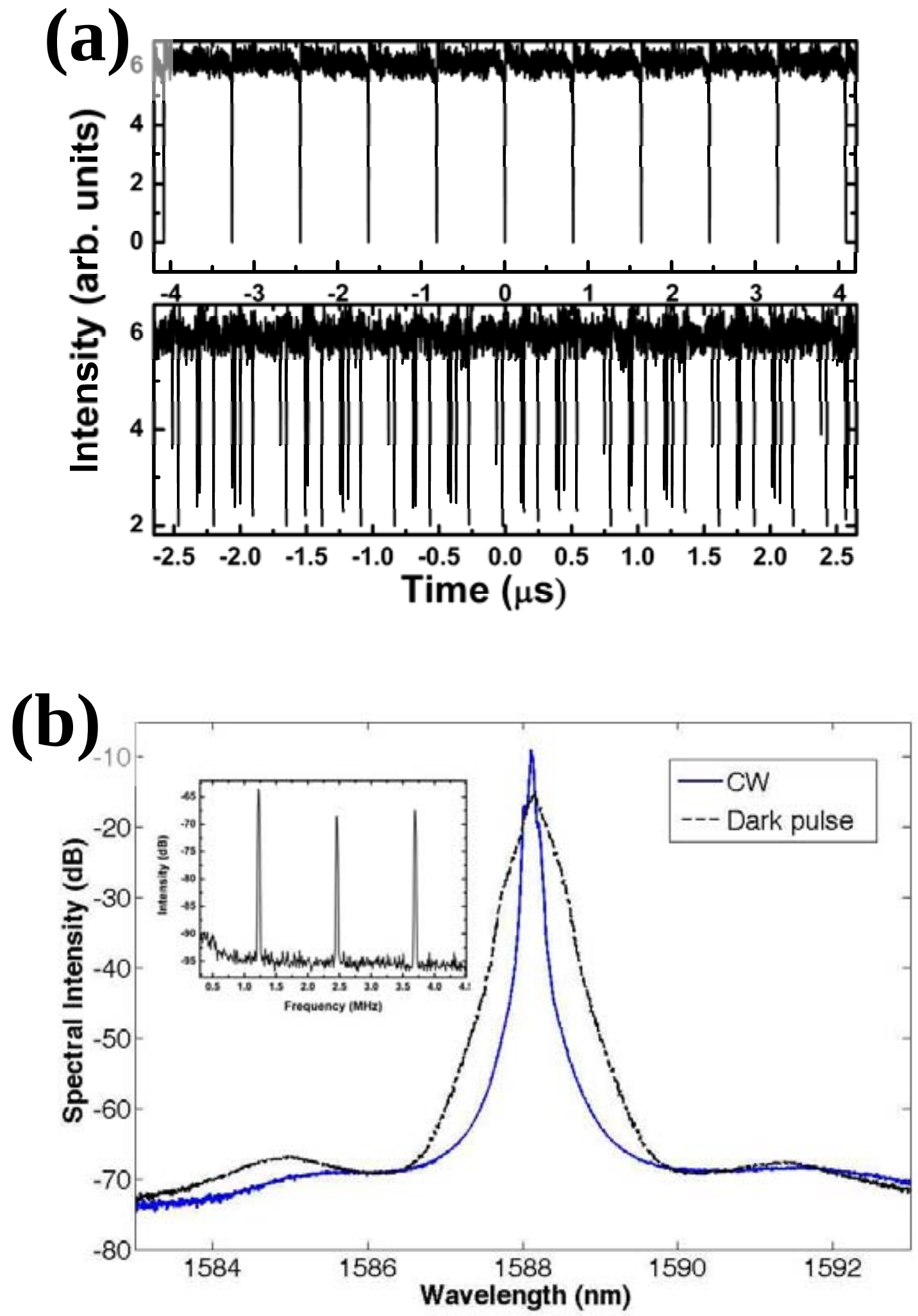

Fig.2 H. Zhang et al

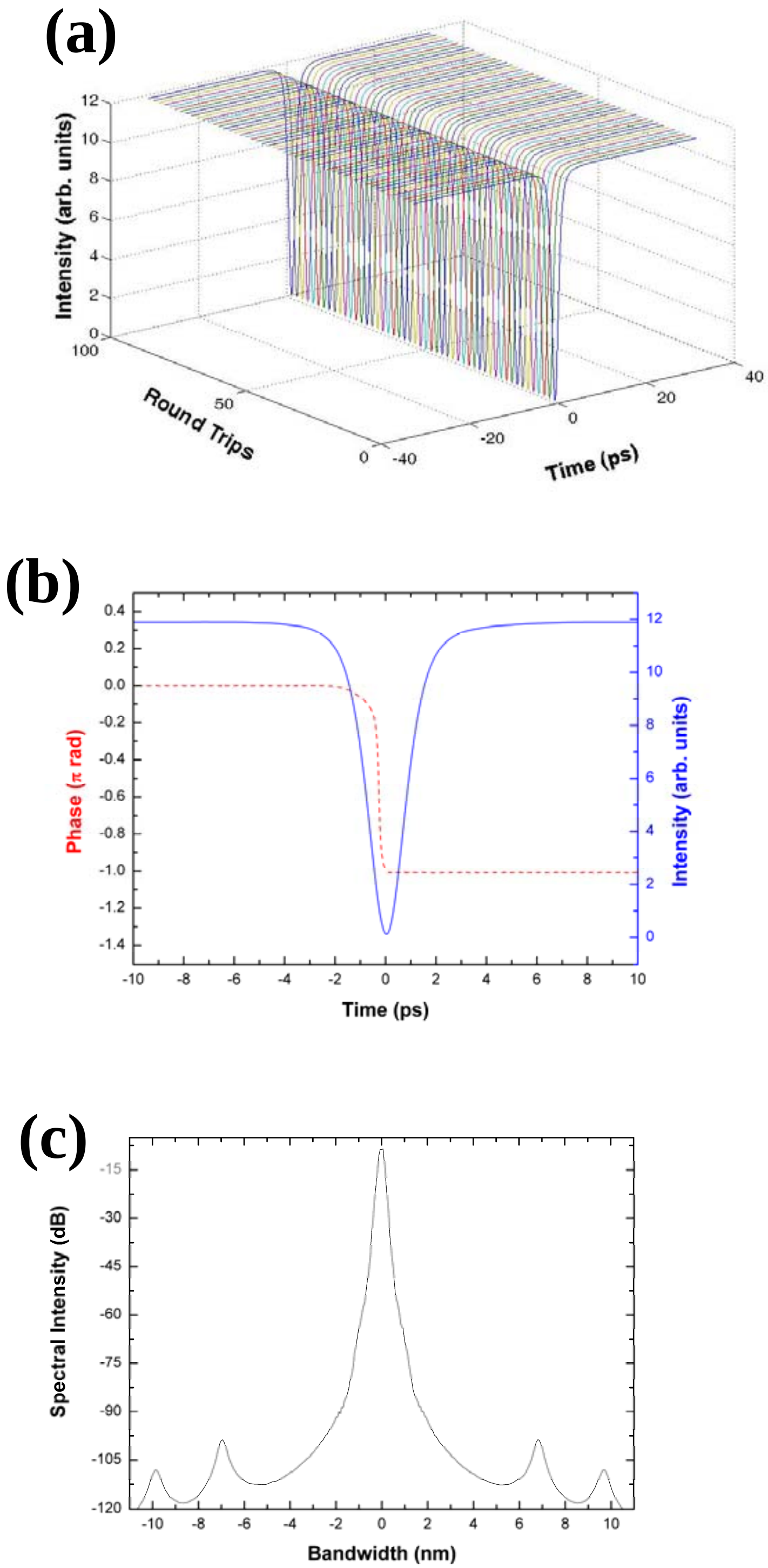

Fig.3 H. Zhang et al